\documentclass{elsart}

\usepackage{natbib,amsmath,graphicx,epsfig,epstopdf,lineno}
\usepackage{color}

 \definecolor{darkblue}{rgb}{0.0, 0.0, 0.55}

\usepackage{setspace}
\usepackage{amssymb,amsfonts,amsmath}
\journal{arXiv.org}

\begin{document}
\begin{frontmatter}

\title{Space periodic Jacobi elliptic solution for triad modified Schr\"odinger equations }

\author{Usama Kadri}
\ead{ukadri@mit.edu}
\address{Department of Mechanical Engineering, Massachusetts Institute of Technology, Cambridge, MA 02139, USA}

\begin{abstract}
We present an analytical solution for triad nonlinear evolution equations with modified Schr\"odinger terms. An example for application in compressible water waves is presented. 
\end{abstract}
\end{frontmatter}
\section{Introduction}
We consider a two-dimensional problem of an interacting wave triad of the form
\small
\begin{equation} \label{eq:schrodinger}
\Psi_{j,t}(x,t)=i\alpha_j\left[\Psi_{j,xx}(x,t)+\delta_j^2\Psi_{j}(x,t)\right]+\gamma_jV_{j}(t)\Psi_{j}(x,t) \qquad j=1,2,3
\end{equation}
\normalsize
with $\alpha_j$, $\delta_j$, and $\gamma$ are parameters of the physical problem. We also assume that the relations
\begin{equation} \label{eq:relations}
V_{1}\Psi_{1}=\Psi_2\Psi_3, \quad V_{2}\Psi_{2}=\Psi_1\Psi_3^*, \quad V_{3}\Psi_{3}=\Psi_1\Psi_2^*,
\end{equation}
are satisfied, $\delta_2=\delta_3=\delta_1/2\equiv\delta/2$, and asterisks denote complex conjugates. The objective is to derive a periodic analytical solution, for application in long and short wave triad interactions.

\section{Solution}
Redefine $\Psi_j(x,t)=g_j(x)f_j(t)$, so that equations (\ref{eq:schrodinger}) can be written as
\begin{equation} \label{eq:g1}
g_1f_{1,t}=i\alpha_1\left[g_{1,xx}+\delta_1^2g_{1}\right]f_1+\gamma_1g_2g_3f_2f_3 
\end{equation}
\begin{equation} \label{eq:g2}
g_2f_{2,t}=i\alpha_2\left[g_{2,xx}+\delta_2^2g_{2}\right]f_2+\gamma_2g_1g_3^*f_1f_3^*
\end{equation}
\begin{equation} \label{eq:g3}
g_3f_{3,t}=i\alpha_3\left[g_{3,xx}+\delta_3^2g_{3}\right]f_3+\gamma_3g_1g_2^*f_1f_2^*
\end{equation}

Based on (\ref{eq:relations}), $g_1=g_2g_3$, $g_2=g_1g_3^*$, and $g_3=g_1g_2^*$, so that  equations (\ref{eq:g1}), (\ref{eq:g2}), and (\ref{eq:g3}), can be rewritten as
\begin{equation} \label{eq:g1b}
f_{1,t}=\gamma_1f_2f_3 +i\alpha_1\left[g_{1,xx}+\delta_1^2g_{1}\right]f_1g_1^{-1}
\end{equation}
\begin{equation} \label{eq:g2b}
f_{2,t}=\gamma_2f_1f_3^*+i\alpha_2\left[g_{2,xx}+\delta_2^2g_{2}\right]f_2g_2^{-1}
\end{equation}
\begin{equation} \label{eq:g3b}
f_{3,t}=\gamma_3f_1f_2^*+i\alpha_3\left[g_{3,xx}+\delta_3^2g_{3}\right]f_3g_3^{-1}
\end{equation}

Now we can seek a solution in two parts. The first part requires that 
\begin{equation} \label{eq:g}
g_{j,xx}+\delta_j^2g_{j}=0, \qquad j=1,2,3.
\end{equation}

A general solution of (\ref{eq:g}) is given by
\begin{equation}
g_j=a_je^{i\delta_jx}+b_je^{-i\delta_jx}.
\end{equation}

For the second part of the solution we need to solve the following simplified system of three ordinary differential equations that amplitudes satisfy
\begin{equation} \label{eq:f}
f_{1,t}=\gamma_1f_2f_3, \qquad f_{2,t}=\gamma_2f_1f_3^*, \qquad
f_{3,t}=\gamma_3f_1f_2^* 
\end{equation}

Multiplying (\ref{eq:f}) by $f_j^*$ and adding its conjugate multiplied by $f_j$, for j=1,2,3 respectively, we obtain the following set of equations
\begin{equation} \label{eq:f1_a}
f_1^*f_{1,t}+f_1f_{1,t}^* = \gamma_1\left(f_1^*f_2f_3+f_1f_2^*f_3^*\right)
 \end{equation}
\begin{equation} \label{eq:f1_a}
  f_2^*f_{2,t}+f_2f_{2,t}^* = \gamma_2\left(f_1^*f_2f_3+f_1f_2^*f_3^*\right)
 \end{equation}
 \begin{equation} \label{eq:f1_a}
 f_3^*f_{3,t}+f_3f_{3,t}^* = \gamma_3\left(f_1^*f_2f_3+f_1f_2^*f_3^*\right)
 \end{equation}

More compactly we can write $|f_j|_{,t}^2 = 2\gamma_j \Im\{f_1^*f_2f_3\}$, where the Hamiltonian $\Im\{f_1^*f_2f_3\}$ is a constant of the motion (\cite{holm2002stepwise}). Now define $\mathbb{Z}_{,t}=\Im\{f_1^*f_2f_3\}$ gives $|f_j|^2 = 2\gamma_j \mathbb{Z} + \psi_{0j}^2$. In order to carry on with the solution the signs of $\gamma_j$ have to be determined. Note that for a resonating triad, $\gamma_1+\gamma_2+\gamma_3=0$, (\cite{lynch2003resonant}), thus one has a different sign than the others. Assume, with no loss of generality, that $\gamma_1$ is negative, that $\Psi_1(x,t=0)=\psi_{01}=0$, and that $|\psi_{03}|<|\psi_{02}|$ we obtain
%
%
\begin{equation} \label{eq:Z}
\mathbb{Z}_{,t}=\sqrt{-8|\gamma_1|\gamma_2\gamma_3\mathbb{Z}\left(\mathbb{Z}+\frac{\psi_{02}^2}{\gamma_2}\right)\left(\mathbb{Z}+\frac{\psi_{03}^2}{\gamma_3}\right)}
\end{equation}

This is an elliptic function with a solution given by (see \cite{byrd1971handbook}, equation 236.00, p.79)
\begin{equation} \label{eq:Z_final}
\mathbb{Z}=-\frac{\psi_{03}^2}{\gamma_3}\textrm{sn}^2(u,k)
\end{equation}
where $\textrm{sn}(u,k)$ is the sine amplitude Jacobian elliptic function of argument $u$, and modulus $k$ given by
\begin{equation} \label{eq:u_k}
u=\sqrt{2|\gamma_1|\gamma_3}|\psi_{02}| t, \qquad k=\frac{|\psi_{03}|}{|\psi_{02}|}\sqrt{\frac{\gamma_2}{\gamma_3}}
\end{equation}
and the expression for $|f_j|$ are 
\begin{equation} \label{eq:f1_final}
|f_j|^2=|\psi_{0j}|^2-2\gamma_j\frac{|\psi_{03}|^2}{\gamma_3}\textrm{sn}^2(u,k)
\end{equation}

Finally, the analytical solution is given by
\begin{equation} \label{eq:f1_final}
|\Psi_1(x,t)|^2=-2\gamma_1\frac{|\psi_{03}|^2}{\gamma_3}\textrm{sn}^2(u,k) \left[\exp{(2i\delta x)}+\exp{(-2i\delta x})\right]
\end{equation}
\begin{equation} \label{eq:f2_final}
|\Psi_2(x,t)|^2=\left[|\psi_{02}|^2-2\gamma_2\frac{|\psi_{03}|^2}{\gamma_3}\textrm{sn}^2(u,k)\right] \left[\exp{(i\delta x)}+\exp{(-i\delta x})\right]
\end{equation}
\begin{equation} \label{eq:f3_final}
|\Psi_3(x,t)|^2=|\psi_{03}|^2\left[1-2\textrm{sn}^2(u,k)\right] \left[\exp{(i\delta x)}+\exp{(-i\delta x})\right]
\end{equation}

\section{Application}
The solution presented here can be applied in various long-short wave triad interactions, such as Rossby-type waves (see \cite{pedlosky1987geophysical}; \cite{charney1948scale}), or wave motion in an inhomogeneous plasma (\cite{hasegawa1977pseudo}). Nevertheless, the following example considers the interaction of two surface gravity waves with an acoustic wave in a mechanism similar to that proposed by \cite{longuet1950theory}, and more recently by \cite{kadri2013generation} and \cite{kadri2015wave}. 

Given an acoustic wave and two gravity waves with potential amplitudes $\phi_{a}$, $\phi_{g1}$, and $\phi_{g2}$, satisfying the following evolution equations
 \begin{equation}
\phi_{a,t} = -\frac{ic^2\delta}{2\omega h}\left(\phi_{a,xx}+\frac{4\omega^2}{c^2}\phi_a\right)-\frac{2\omega}{hc}\phi_{g1}\phi_{g2} 
\end{equation}
 \begin {equation}
 \phi_{g1,t} = \frac{2\omega^3}{gc}\phi_{a}\phi_{g2}^*; \quad  \phi_{g2,t} = \frac{2\omega^3}{gc}\phi_{a}\phi_{g1}^*
 \end{equation}
where $c=1500$ m/s, is the speed of sound in water, $\omega$ is the frequency of the gravity waves, $h$ is the water depth. The solution of evolution equations is then given by
\begin{equation} \label{eq:f1_final}
|\Phi_a(x,t)|^2=\frac{2g |\phi_{0(g2)}|^2}{\omega h}\textrm{sn}^2\left(u,k\right) \left(e^{4i \omega x/c}+e^{-4i\omega x/c}\right)
\end{equation}
\begin{equation} \label{eq:f2_final}
|\Phi_{g1}(x,t)|^2=\left[|\phi_{0(g1)}|^2-2|\phi_{0(g2)}|^2\textrm{sn}^2(u,k)\right] \left(e^{2i\omega x/c}+e^{-2i\omega x/c}\right)
\end{equation}
\begin{equation} \label{eq:f3_final}
|\Phi_{g2}(x,t)|^2=|\phi_{0(g2)}|^2\left[1-2\textrm{sn}^2(u,k)\right] \left(e^{2i\omega x/c}+e^{-2i\omega x/c}\right)
\end{equation}
with $u=2\sqrt{2/gh}\omega^2/c$, and $k=|\phi_{0(g2)}|/|\phi_{0(g1)}|$. 
%
%
 


\bibliographystyle{elsart-harv}

\begin{thebibliography}{}


\bibitem[\textit{Byrd and Friedman}(1971)]{byrd1971handbook}
Byrd, P. F. and Friedman M. D. (1971), Handbook of elliptic integrals for engineers and scientists,{\it Springer-Verlag}, pp. 358.

\bibitem[\textit{Charney}(1948)]{charney1948scale}
Charney, J.G. (1948) On the Scale of Atmospheric Motions. {\it Geofys. Publ.}, 17, No. 2, 17pp.

\bibitem[\textit{Hasegawa and Mima}(1977)]{hasegawa1977pseudo}
Hasegawa, A. and Mima, K. (1977) Pseudo-three-dimensional turbulence in magnetized nonuniform plasmas.
{\it Phys. Fluids}, 21(1), 87Ð92.

\bibitem[\textit{Holm and Lynch}(2002)]{holm2002stepwise}
Holm, D.D. and Lynch, P., (2002) Stepwise Precession of the Resonant Swinging Spring. {\it SIAM J. Appl. Dy- nam. Systems}, 1, 44Ð64.

\bibitem[\textit{Kadri}(2015)]{kadri2015wave}
Kadri, U. (2015) Wave motion in a heavy compressible fluid: revisited, {\it European Journal of Mechanics -B/Fluids}, \textbf{49(A)}, 50--57, 10.1016/j.euromechflu.2014.07.008.

\bibitem[\textit{Kadri and Stiassnie}(2013)]{kadri2013generation}
Kadri, U. and Stiassnie, M. (2013) Generation of an acoustic-gravity wave by two gravity waves, and their mutual interaction. {\it J. Fluid Mech.}, \textbf{735} R6, doi:10.1017/jfm.2013.539.

\bibitem[\textit{Longuet--Higgins}(1950)]{longuet1950theory}
Longuet--Higgins, M. S. (1950) A theory of the origin of microseisms,{\it Philos. Trans. R. Soc. London, Ser. A}, \textbf{243}, 1--35, doi:10.1098/rsta.1950.0012.

\bibitem[\textit{Pedlosky}(1987)]{pedlosky1987geophysical}
Pedlosky, J. (1987) Geophysical Fluid Dynamics. Second edition, {\it Springer}, New York. 710pp.

\bibitem[\textit{Lynch et al.}(2003)]{lynch2003resonant}
Lynch, P., Eireann, M., Hill, H. (2003) Resonant Rossby Wave Triads and the Swinging Spring, {\it Bull. Amer. Met. Soc.}, DOI: 10.1175/BAMS-84-5-605.

%
%
%
%
%

\end{thebibliography}

\end{document}